\documentclass{amsart}%
\usepackage{amsfonts}
\usepackage{amsmath}
\usepackage{amsxtra}
\usepackage{amssymb}
\usepackage{graphicx}%
\providecommand{\U}[1]{\protect\rule{.1in}{.1in}}
\providecommand{\U}[1]{\protect\rule{.1in}{.1in}}
\providecommand{\U}[1]{\protect\rule{.1in}{.1in}}
\providecommand{\U}[1]{\protect\rule{.1in}{.1in}}
\theoremstyle{plain}
\newtheorem{theorem}{Theorem}

\begin{document}
\title{Unstable Galaxy Models}
\author{Zhiyu Wang}
\address{School of Mathematical Sciences\\
Peking University\\
Beijing, 100871, P. R. China}
\author{Yan Guo}
\address{Division of Applied Mathematics \\
Brown University\\
Providence, RI 02912, USA}
\author{Zhiwu Lin}
\address{School of Mathematics\\
Georgia Institute of Technology\\
Atlanta, GA 30332, USA}
\author{Pingwen Zhang}
\address{School of Mathematical Sciences\\
Peking University\\
Beijing, 100871, P. R. China}

\begin{abstract}
The dynamics of collisionless galaxy can be described by the Vlasov-Poisson
system. By the Jean's theorem, all the spherically symmetric steady galaxy
models are given by\ a distribution of $\Phi(E,L)$, where $E$ is the particle
energy and $L$ the angular momentum. In a celebrated Doremus-Feix-Baumann
Theorem \cite{dbf73}, the galaxy model $\Phi(E,L)$ is stable if the
distribution $\Phi$ is monotonically decreasing with respect to the particle
energy $E.$ On the other hand, the stability of $\Phi(E,L)$ remains largely
open otherwise. Based on a recent abstract instability criterion of Guo-Lin
\cite{GL}, we constuct examples of unstable galaxy models of $f(E,L)$ and
$f\left(  E\right)  \ $in which $f$ fails to be monotone in $E.$

\end{abstract}
\maketitle

\section{\textbf{Introduction}}

A galaxy is an ensemble of billions of stars, which interact by the
gravitational field they create collectively. For galaxies, the collisional
relaxation time is much longer than the age of the universe (\cite{BT}). The
collisions can therefore be ignored and the galactic dynamics is well
described by the Vlasov - Poisson system (collisionless Boltzmann equation)
\begin{equation}
\partial_{t}f+v\cdot\nabla_{x}f-\nabla_{x}U_{f}\cdot\nabla_{v}f=0,\text{
\ \ \ \ \ }\Delta U_{f}=4\pi\int_{\mathbf{R}^{3}}f(t,x,v)dv, \label{vp}%
\end{equation}
where $\left(  x,v\right)  \in\mathbb{R}^{3}\times\mathbb{R}^{3}$, $f(t,x,v)$
is the distribution function and $U_{f}\left(  t,x\right)  $ is its
self-consistent gravitational potential. The Vlasov-Poisson system can also be
used to describe the dynamics of globular clusters over their period of
orbital revolutions (\cite{fp84}). One of the central questions in such
galactic problems, which has attracted considerable attention in the
astrophysics literature, of \cite{bertin}, \cite{BT}, \cite{fp84},
\cite{merritt} and the references there, is to determine \textit{dynamical
stability }of steady galaxy models. Stability study can be used to test a
proposed configuration as a model for a real stellar system. On the other
hand, instabilities of steady galaxy models can be used to explain some of the
striking irregularities of galaxies, such as spiral arms as arising from the
instability of an initially featureless galaxy disk (\cite{bertin}),
(\cite{palmer}).

In this article, we consider stability of spherical galaxies, which are the
simplest elliptical galaxy models. Though most elliptical galaxies are known
to be non-spherical, the study of instability and dynamical evolution of
spherical galaxies could be useful to understand more complicated and
practical galaxy models. By Jeans's Theorem, a steady spherical galaxy is of
the form $f(x,v)\equiv\mu(E,L)$, where the particle energy and total momentum
are $E=\frac{1}{2}|v|^{2}+U(x),\ L=\left\vert x\times v\right\vert ,$ and
$U_{\mu}(x)=U\left(  \left\vert x\right\vert \right)  $ satisfies the
self-consistent nonlinear Poisson equation%
\[
\Delta U=4\pi\int_{\mathbf{R}^{3}}\mu(E,L)dv
\]
The isotropic models take the form $f(x,v)\equiv\mu(E).$ The case when
$\mu^{\prime}(E)<0$ (on the support of $\mu(E)$)$\ $has been widely studied
and these models are known to be stable (\cite{an61}, \cite{ant62},
\cite{dbf73}, \cite{ant62}, \cite{dbf76}, \cite{sygnet84}, \cite{KS85},
\cite{aly96}, \cite{gr-king}). To understand such a stability, we expand the
well-known Casimir-Energy functional (as a Liapunov functional)
\begin{equation}
\mathcal{H}\equiv\int\int Q(f)+\frac{1}{2}\int\int|v|^{2}f-\frac{1}{8\pi}%
\int|\nabla_{x}U|^{2} \label{casimir}%
\end{equation}
which is conserved for all time with $\ $%
\begin{equation}
Q^{\prime}(\mu(E))\equiv-E \label{monotone}%
\end{equation}
$($this is possible only if $\mu^{\prime}<0!$). Upon using Taylor expansion
around a steady galaxy model of $[f(E,L),U],$ the first order variation
vanishes due to the choice of (\ref{monotone}), and the second order variation
takes the form of
\begin{equation}
\mathcal{H}_{f}^{\prime\prime}[g,g]\equiv\frac{1}{2}\int\int_{f>0}\frac{g^{2}%
}{-f^{\prime}(E)}-\frac{1}{8\pi}\int|\nabla_{x}U_{g}|^{2}. \label{h2}%
\end{equation}
The remarkable feature for stability lies in the fact that $\mathcal{H}_{\mu
}^{\prime\prime}[g,g]>0$ if $\mu^{\prime}<0$ (\cite{an61}, \cite{ant62},
\cite{dbf73}, \cite{KS85}). This crucial observation leads to the conclusion
that galaxy models with monotonically decreasing energy $\mu^{\prime}<0$ are
linearly stable. On the other hand, for the case $\mu$ is \textit{not}
monotone in $E$, (\ref{monotone}) breaks down, and $\mathcal{H}_{\mu}%
^{\prime\prime}$ is not well-defined, which indicates possible formation of
instability. It is important to note that a negative direction of the
quadratic form $\mathcal{H}_{f}^{\prime\prime}[g,g]<0$ does \textit{not} imply
instability. In \cite{henon73}, an oscillatory instability was found for
certain generalized polytropes $f\left(  E,L\right)  =f_{0}L^{-2m}\left(
E_{0}-E\right)  ^{n-\frac{3}{2}}$ with $n<\frac{3}{2},m<0,\ $by using N-body
code. This instability was later reanalyzed by more sophisticated N-body code
in \cite{bgh86}. Despite progresses made over the years (e.g., \cite{bgh86},
\cite{goodman88}, \cite{merritt}, \cite{palmer}), \textit{no} explicit example
of isotropic galaxy model $\mu(E)$ are known to be unstable.

The difficulty of finding instability lies in the complexity of the linearized
Vlasov-Poisson system around a spatially non-homogeneous $\mu(E,L):$%
\begin{equation}
\partial_{t}g+v\cdot\nabla_{x}g-\nabla_{x}U_{g}\cdot\nabla_{v}\mu-\nabla
_{x}U\cdot\nabla_{v}g=0,\ \ \Delta U_{g}=4\pi\int_{\mathbf{R}^{3}}g(x,v)dv
\label{linearvp}%
\end{equation}
for which the construction for dispersion relation for a growing mode is
mathematically very challenging. In a recent paper (\cite{GL}), a sufficient
condition was derived rigorously as follows:

\begin{theorem}
\label{theorem-insta}Let $[f(E,L),U]$ be a steady galaxy model. Assume that
$f(E,L)$ has a compact support in $x$ and $v,$ and $\mu^{\prime}$ is bounded.
Define auxiliary quadratic form for a spherically symmetric function
$\phi(|x|)$ as
\begin{equation}
\lbrack A_{0}\phi,\phi]\equiv\int_{\mathbf{R}^{3}}|\nabla\phi|^{2}dx+32\pi
^{3}\int f^{\prime}(E,L)\int_{r_{1}(E,L)}^{r_{2}(E,L)}\left(  \phi-\bar{\phi
}\right)  ^{2}\frac{2LdrdEdL}{\sqrt{2(E-U(r)-L^{2}/2r^{2})}},
\label{formula-qudratic-A0}%
\end{equation}
where $r_{1}(E,L)$ and $r_{2}(E,L)$ are two distinct roots to the equation
\begin{equation}
E-U(r)-L^{2}/2r^{2}=0 \label{eq}%
\end{equation}
and the average $\bar{\phi}$ is defined as
\[
\bar{\phi}(E,L)=\frac{\int_{r_{1}(E,L)}^{r_{2}(E,L)}\frac{\phi(r)dr}%
{\sqrt{2(E-U_{0}(r)-L^{2}/2r^{2})}}}{\int_{r_{1}(E,L)}^{r_{2}(E,L)}\frac
{dr}{\sqrt{2(E-U_{0}(r)-L^{2}/2r^{2})}}}.
\]
If there exists $\phi(|x|)$ such that $[A_{0}\phi,\phi]<0$, then there exists
an exponentially growing mode to the linearized Vlasov-Poisson system
(\ref{linearvp}).
\end{theorem}

The quadratic form $[A_{0}\phi,\phi]$ in the above instability criterion
involves delicate integrals along particle paths$.$ The purpose of the current
paper is to use numerical computations to construct two explicit examples of
galaxy models for which a test function $\phi\ $satisfying $[A_{0}\phi
,\phi]<0$ exists. This ensures their radial instability by Guo-Lin's Theorem
\ref{theorem-insta}. The first example is an anisotropic model with radial
instability. There are two differences to the oscillatory instability found in
\cite{henon73} and \cite{bgh86}: here the distribution function is
non-singular and the instability is non-oscillatory. The second example is a
non-singular isotropic model with radial instability. To our knowledge, this
provides the first example of unstable isotropic model.

Even though the galaxy models studied here are not actual ones observed, our
results demonstrates how to apply Guo-Lin's Theorem \ref{theorem-insta} to
detect possible instability for a given galaxy model. It is our hope to foster
interactions between mathematical and astronomical communities and to advance
the study of instability of real galaxy models.

\section{Examples of Unstable galaxy models}

\begin{figure}[tbh]
\centering
\includegraphics[scale=0.4]{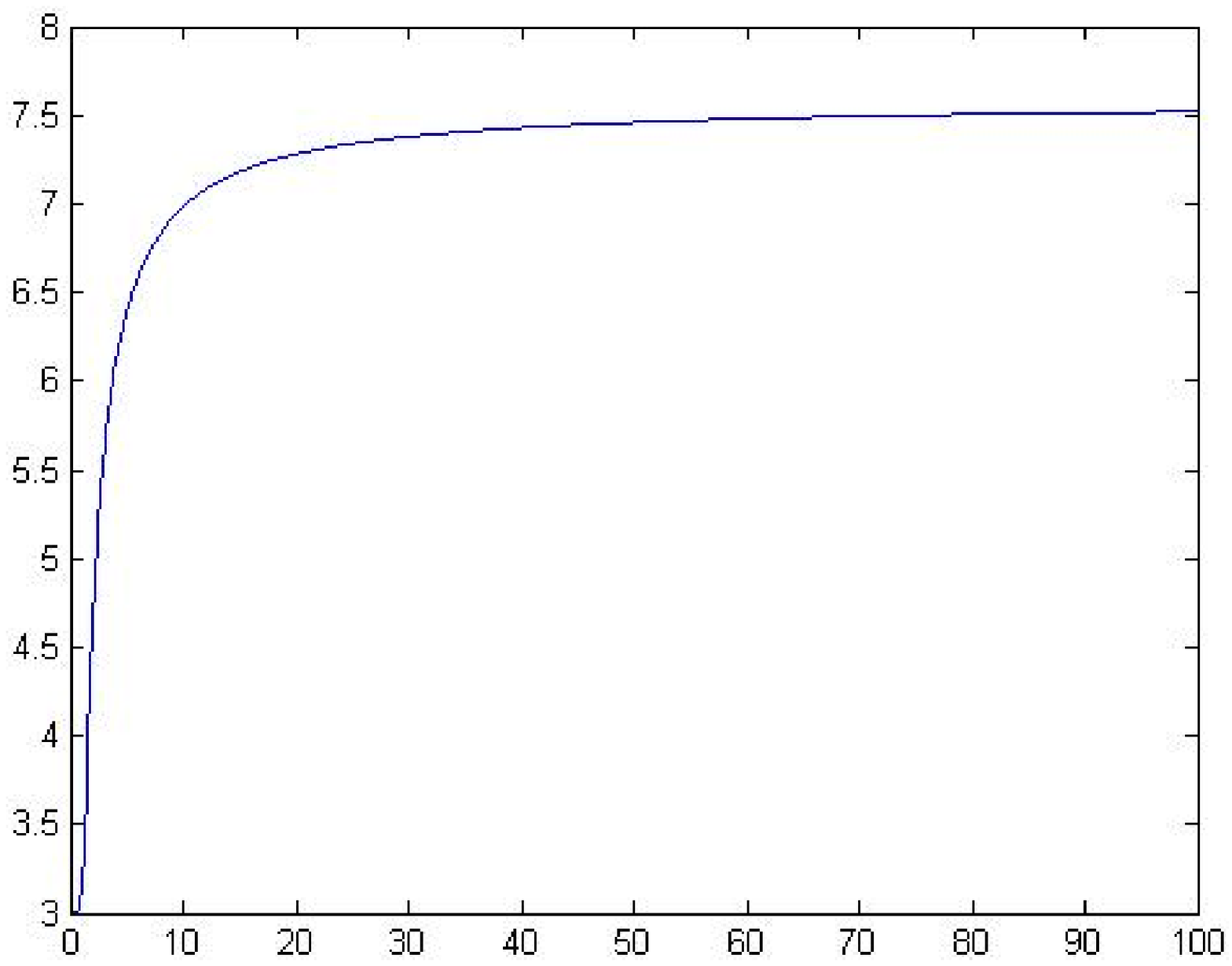} \caption{$U(r)$}%
\label{potential}%
\end{figure}

\textbf{Example} 1. \textbf{Unstable Galaxy Model Depending on }%
$E~$\textbf{and} $L.$ We define the distribution function $f_{0}\left(
E,L\right)  =\mu(E)L^{4}$ where
\begin{equation}
\mu(E)=\left\{
\begin{array}
[c]{l}%
0\quad E<4\\
2.25(E-4)^{2}\quad4\leq E\leq4.4\\
(5-E)^{2}\quad4.4\leq E\leq5\\
0\quad E>5
\end{array}
\right.  \label{mu1}%
\end{equation}
The graph of $\mu(E)$ is showed in Figure \ref{phi} below. \begin{figure}[tbh]
\centering
\includegraphics[scale=0.4]{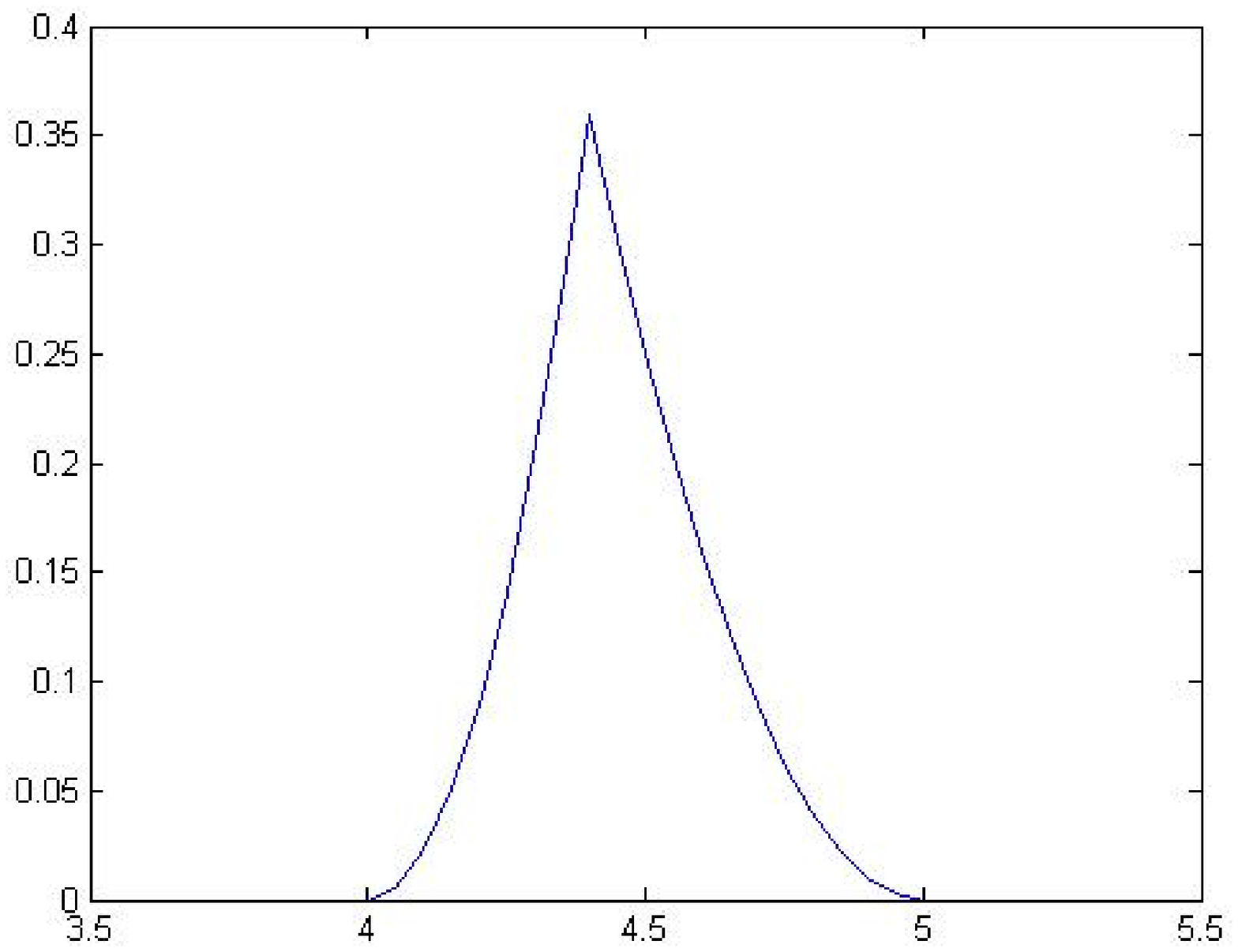} \caption{$\mu(E)$}%
\label{phi}%
\end{figure}Choosing $U(0)=3$, we numerically calculate $U(r)$, and the graph
of $U(r)$ is showed in Figure \ref{potential}. By choosing the test function $\phi=e^{-r},$ then numerical
computation below shows
\[
\lbrack A_{0}\phi,\phi]<\pi-45<0,
\]
and hence $f_{0}$ is unstable by Theorem \ref{theorem-insta}.

\begin{figure}[tbh]
\centering
\includegraphics[scale=0.4]{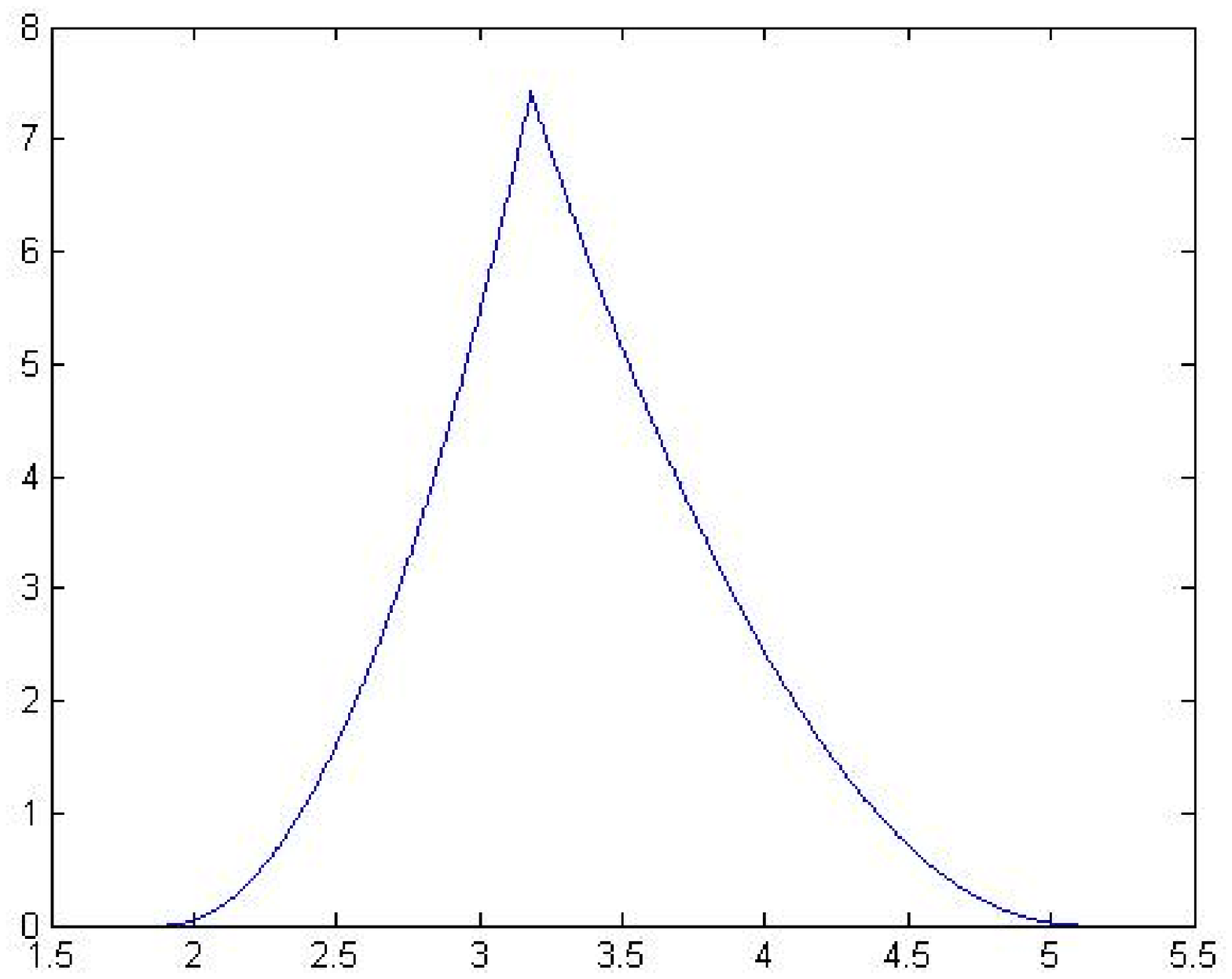} \caption{$\mu(E)$}%
\label{mu}%
\end{figure}

\begin{figure}[tbh]
\centering
\includegraphics[scale=0.4]{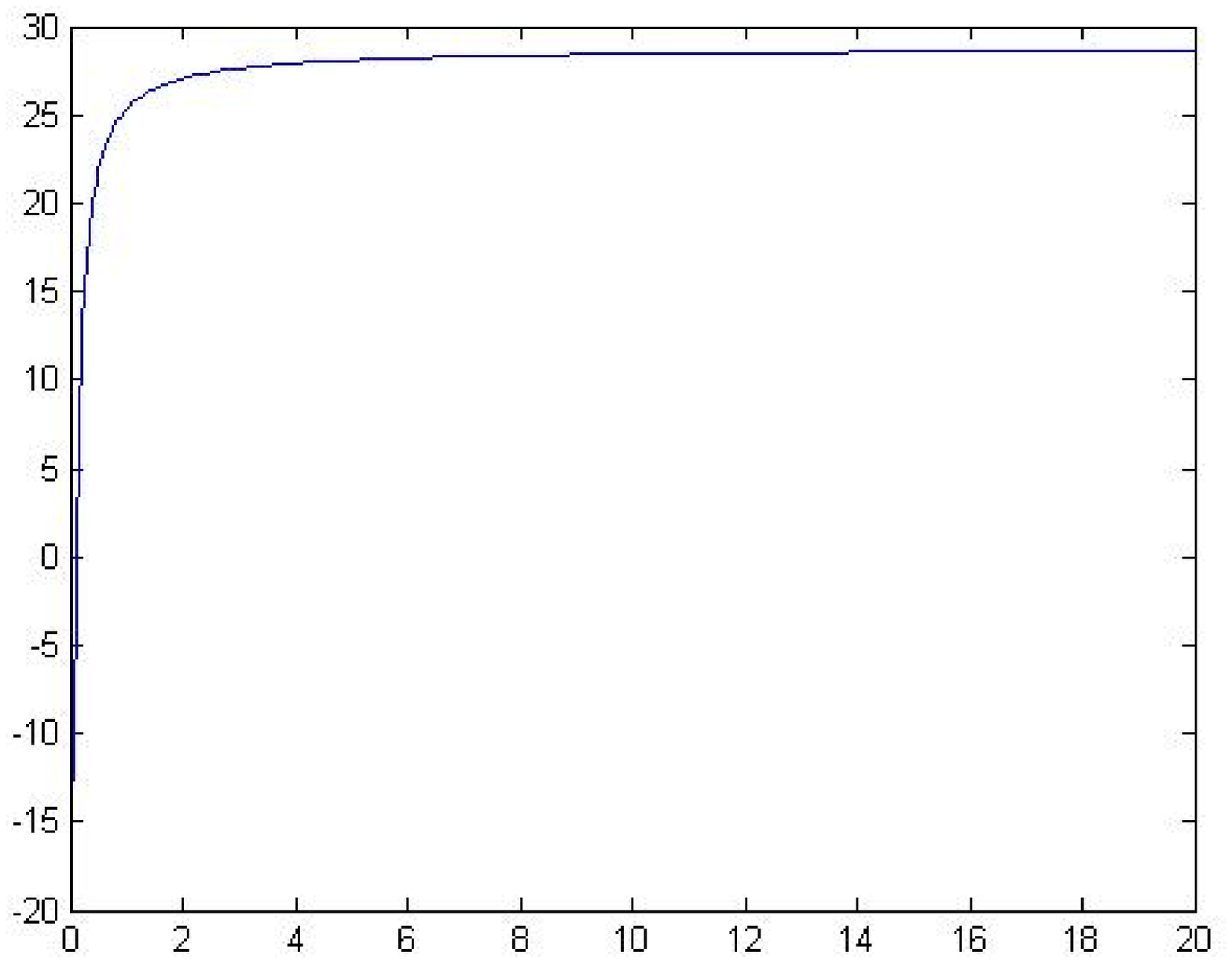} \caption{$U(r)$}%
\label{potential2}%
\end{figure}

\textbf{Example} 2. \textbf{Unstable Galaxy Model Depending Only on }$E.$ Let
$f_{0}(E)=\mu(E)$, where
\begin{equation}
\mu(E)=\left\{
\begin{array}
[c]{l}%
0\quad E<a\\
C_{1}t_{1}^{k}(E-a)^{k}\quad a\leq E\leq E_{1}\\
C_{1}(E_{0}-E)^{k}\quad E_{1}\leq E\leq E_{0}\\
0\quad E>E_{0}%
\end{array}
\right.  \label{mu2}%
\end{equation}
We choose
\[
%\begin{aligned}
C_{1}=2\newline,\ E_{0}=5.1,\ \newline a=1.9\newline,\ k=2.01\newline%
,\ t_{1}=1.5.
%\end{aligned}
\]
Choose $U_{0}=-15.1$, the graphs of $\mu(E)$ and $U$ are shown in Figures
\ref{mu} and \ref{potential2}. \textbf{ }We choose the test function\textbf{ }%
\[
\phi=v_{1}e^{-r}+v_{2}e^{-\frac{1}{2}r}+v_{3}e^{-\frac{1}{3}r}+v_{4}%
e^{-2r}+v_{5}e^{-3r}+v_{6}e^{-4r}+v_{7}e^{-5r},
\]
with\textbf{ }$v_{1},v_{2},...v_{7}$ given by (\ref{vector-v}). Then\textbf{
}$[A_{0}\phi,\phi]=-24.\,\allowbreak016<0,$ and instability of the profile
(\ref{mu2}) follows from Theorem \ref{theorem-insta}.

\section{Numerical implementation}

\subsection{\textbf{Numerical computation for Example} 1}

By the results in \cite{1}, for a steady state with $f_{0}\left(  E,L\right)
=\mu\left(  E\right)  L^{2l}\ $of Vlasov-Poisson system, the steady potential
$U\left(  r\right)  $ satisfies the equation
\begin{equation}
U^{\prime}(r)=\frac{2^{l+7/2}\pi^{2}c_{l,-1/2}}{r^{2}}\int_{0}^{r}%
s^{2l+2}g_{l+1/2}(U(s))ds,\quad r>0 \label{imp}%
\end{equation}
with $g_{m}(u)=\int_{u}^{\infty}\mu(E)(E-u)^{m}dE,\quad u\in(-\infty,\infty)$,
$\ $\ and
\[
c_{a,b}=\int_{0}^{1}s^{a}(1-s)^{b}ds=\frac{\Gamma(a+1)\Gamma(b+1)}%
{\Gamma(a+b+2)},\ \ a>-1,b>-1,
\]
where $\Gamma$ denotes the gamma function. The boundary condition is
$\lim_{r\rightarrow\infty}U(r)=0$.

The computation of finding $\phi$ with $[A_{0}\phi,\phi]<0\ $is carried out in
the following steps.

\textit{Step 1.} \textit{Computation of potential }$U.$ For Example 1,
$f_{0}\left(  E,L\right)  =\mu\left(  E\right)  L^{4}$ where $\mu\left(
E\right)  $ is given by (\ref{mu1}), so $l=2$ and
\[
g_{l+1/2}(u)=\int_{u}^{4.4}2.25(E-4)^{2}(E-u)^{2.5}dE+\int_{4.4}^{5}%
(5-E)^{2}(E-u)^{2.5}dE.
\]
Letting $M=2^{l+7/2}\pi^{2}c_{l,-1/2}$, we obtain from (\ref{imp})
\[
r^{2}U^{\prime\prime}(r)+2rU(r)=Mr^{2l+2}g_{l+1/2}(U(r)),
\]
which is equivalent to
\[
U^{\prime\prime}(r)=-\frac{2}{r}U^{\prime}(r)+Mr^{2l}g_{l+1/2}(U(r)).
\]
We use the boundary conditions $U^{\prime}(0)=0$ and $U(0)=3$. Let
$u_{1}(r)=U(r),u_{2}(r)=u_{1}^{\prime}(r)$, we transform the above equation to
the following system
\begin{equation}
\left\{
\begin{array}
[c]{l}%
u_{1}^{\prime}(r)=u_{2}\\
u_{2}^{\prime}(r)=-\frac{2}{r}u_{2}+Mr^{2l}g_{l+1/2}(u_{1})
\end{array}
\right.  \label{1}%
\end{equation}
with $u_{1}(r)=3$ and $u_{2}(0)=0$. We remark that we can subtract the finite
limit of $U$ at infinity from $U$ and redefine $E_{0}$ accordingly. We apply
Runge-Kutta method to solve equations (\ref{1}), as follows
\[
\left\{
\begin{array}
[c]{l}%
\boldsymbol{y}_{n+1}=\boldsymbol{y}_{n}+\dfrac{h}{6}(\boldsymbol{K}%
_{1}+2\boldsymbol{K}_{2}+2\boldsymbol{K}_{3}+\boldsymbol{K}_{4})\\
\boldsymbol{K}_{1}=\boldsymbol{f}(x_{n},\boldsymbol{y}_{n})\\
\boldsymbol{K}_{2}=\boldsymbol{f}(x_{n}+\frac{h}{2},\boldsymbol{y}_{n}%
+\frac{h}{2}\boldsymbol{K}_{1})\\
\boldsymbol{K}_{3}=\boldsymbol{f}(x_{n}+\frac{h}{2},\boldsymbol{y}_{n}%
+\frac{h}{2}\boldsymbol{K}_{2})\\
\boldsymbol{K}_{4}=\boldsymbol{f}(x_{n}+h,\boldsymbol{y}_{n}+h\boldsymbol{K}%
_{2}).
\end{array}
\right.
\]
First let $h=0.1$ to get the values of $U$ and $U^{\prime}$ at points
$x_{n}=nh$, then we use piecewise cubic Hermite interpolation to get an
approximation of $U\left(  r\right)  $.

\begin{figure}[ptb]
\centering
\includegraphics[scale=0.4]{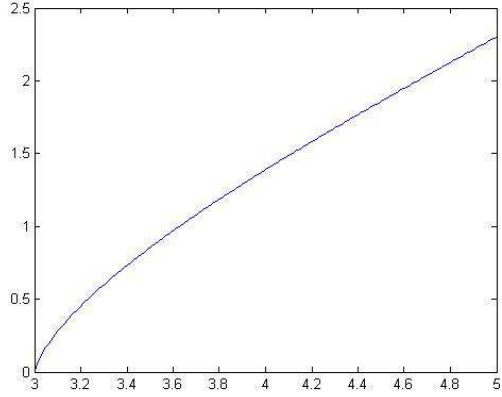} \caption{relationship of $L_{max}$ and
$E$}%
\label{maxL}%
\end{figure}

\textit{Step 2.} \textit{Computation of roots of the equation}
\begin{equation}
E-U(r)-L^{2}/2r^{2}=0. \label{eqn-root}%
\end{equation}
For fixed $E$, this equation has two solutions $r_{1}<r_{2}\ $when $L<L_{\max
}$, one solution $r^{\ast}\ $when $L=L_{\max}$ and no solution when
$L>L_{\max}$. Here, $L_{max}\left(  E\right)  =\sqrt{r^{\ast3}U^{\prime
}(r^{\ast})}$ and $r^{\ast}\left(  E\right)  $ satisfies the equation
\begin{equation}
E-U(r)-\frac{r}{2}U^{\prime}(r)=0, \label{eqn-r*}%
\end{equation}
which comes from the combination of the equations $-U^{\prime}(r)+L^{2}%
/r^{3}=0$ and (\ref{eqn-root}). We employ Newton method to find the unique
root $r^{\ast}\ $of (\ref{eqn-r*}) by the Newton iteration
\[
r_{n+1}=r_{n}-\dfrac{E-U(r)-\frac{r}{2}U^{\prime}(r)}{-\frac{3}{2}U^{\prime
}(r)-\frac{r}{2}U^{\prime\prime}(r)}%
\]
\mbox{Chose an initial point $r_{0}=1$, we get $r^{*}$(with the stopping criterion $|r_{n}-r_{n+1}|<10^{-10}$).}
Table \ref{table1} and Figure \ref{maxL} show the relation of $L_{max}$ to
$E$. \begin{table}[ptb]
\caption{$L_{max}$ for different $E$}%
\label{table1}%
\centering
\begin{tabular}
[c]{c|c}\hline
E & $L_{max}$\\\hline
3.2 & 0.452787235493472\\\hline
3.4 & 0.731586885359610\\\hline
3.6 & 0.969996239842229\\\hline
3.8 & 1.186531828839693\\\hline
4.0 & 1.388987957485868\\\hline
4.2 & 1.581703995145215\\\hline
4.4 & 1.767449993848502\\\hline
4.6 & 1.948250754546466\\\hline
4.8 & 2.125722162034805\\\hline
5.0 & 2.301341325775183\\\hline
\end{tabular}
\end{table}

\begin{figure}[ptb]
\begin{minipage}{0.49\textwidth}
\centering
\includegraphics[scale=0.35]{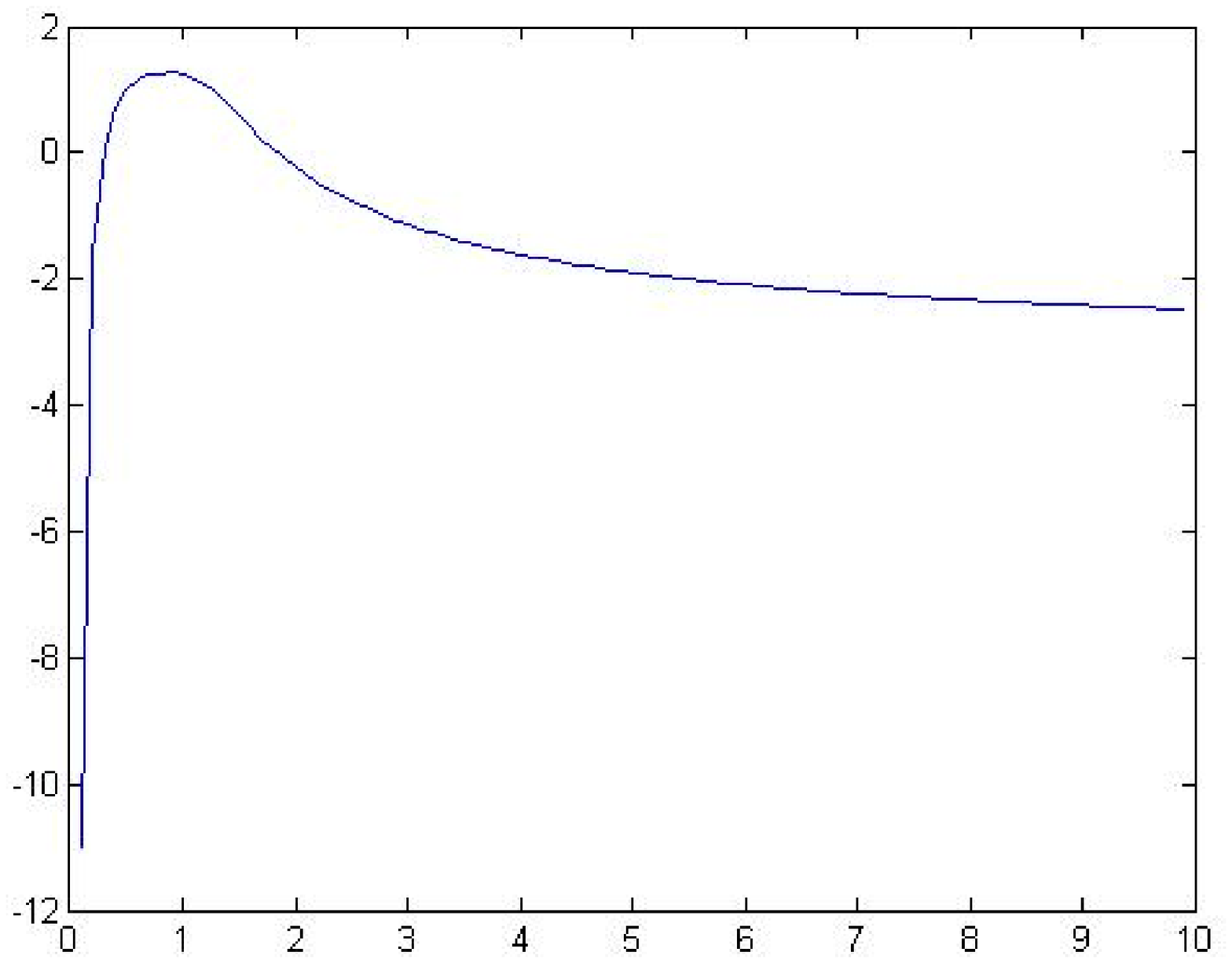}
\caption{$E$=$4.5$,$L$=0.5}\label{E4_5L0_5}
\end{minipage}
\begin{minipage}{0.49\textwidth}
\centering
\includegraphics[scale=0.35]{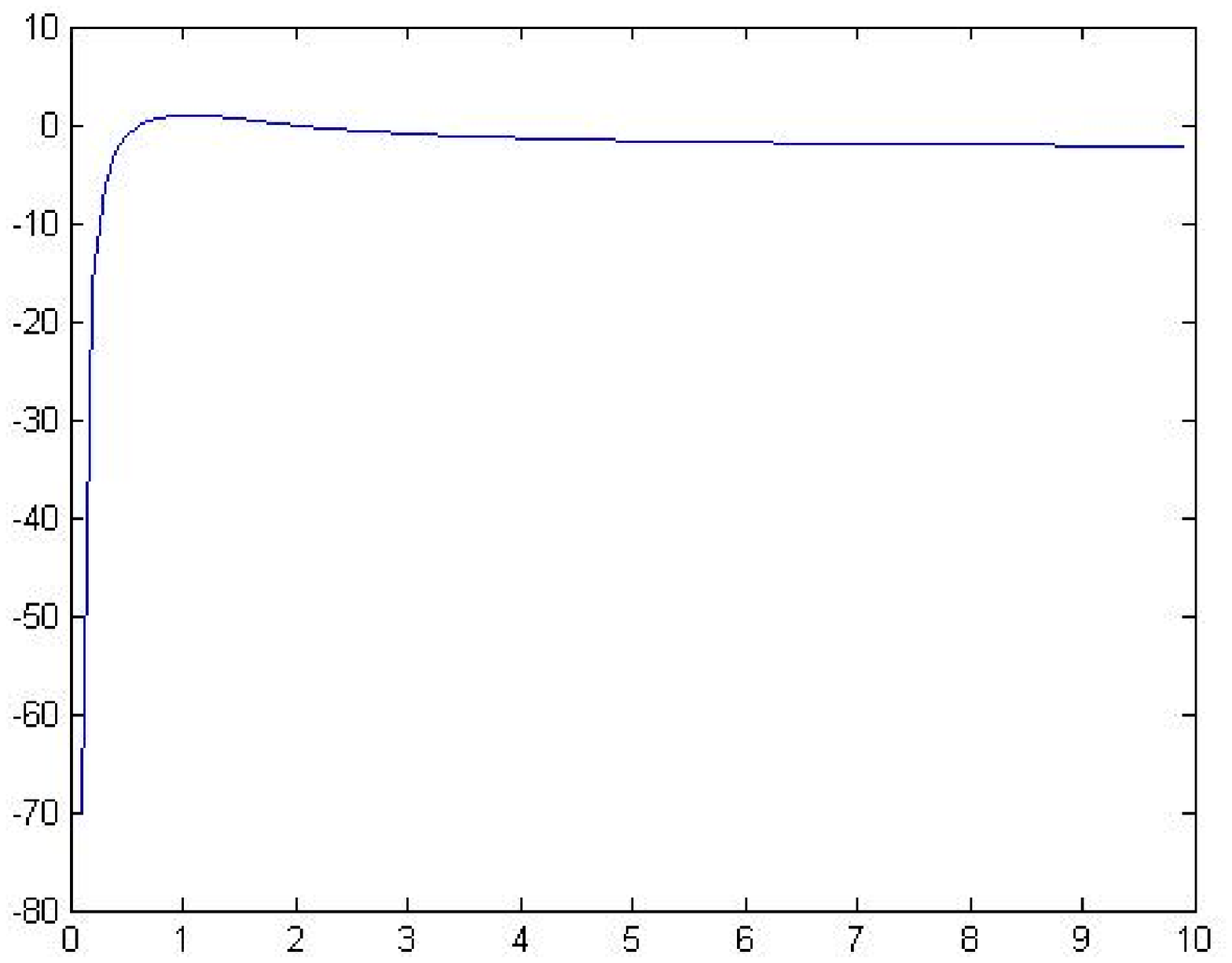}
\caption{$E$=4.8,$L$=1.2}\label{E4_8L1_2}
\end{minipage}
\end{figure}

\begin{figure}[ptb]
\begin{minipage}{0.49\textwidth}
\centering
\includegraphics[scale=0.35]{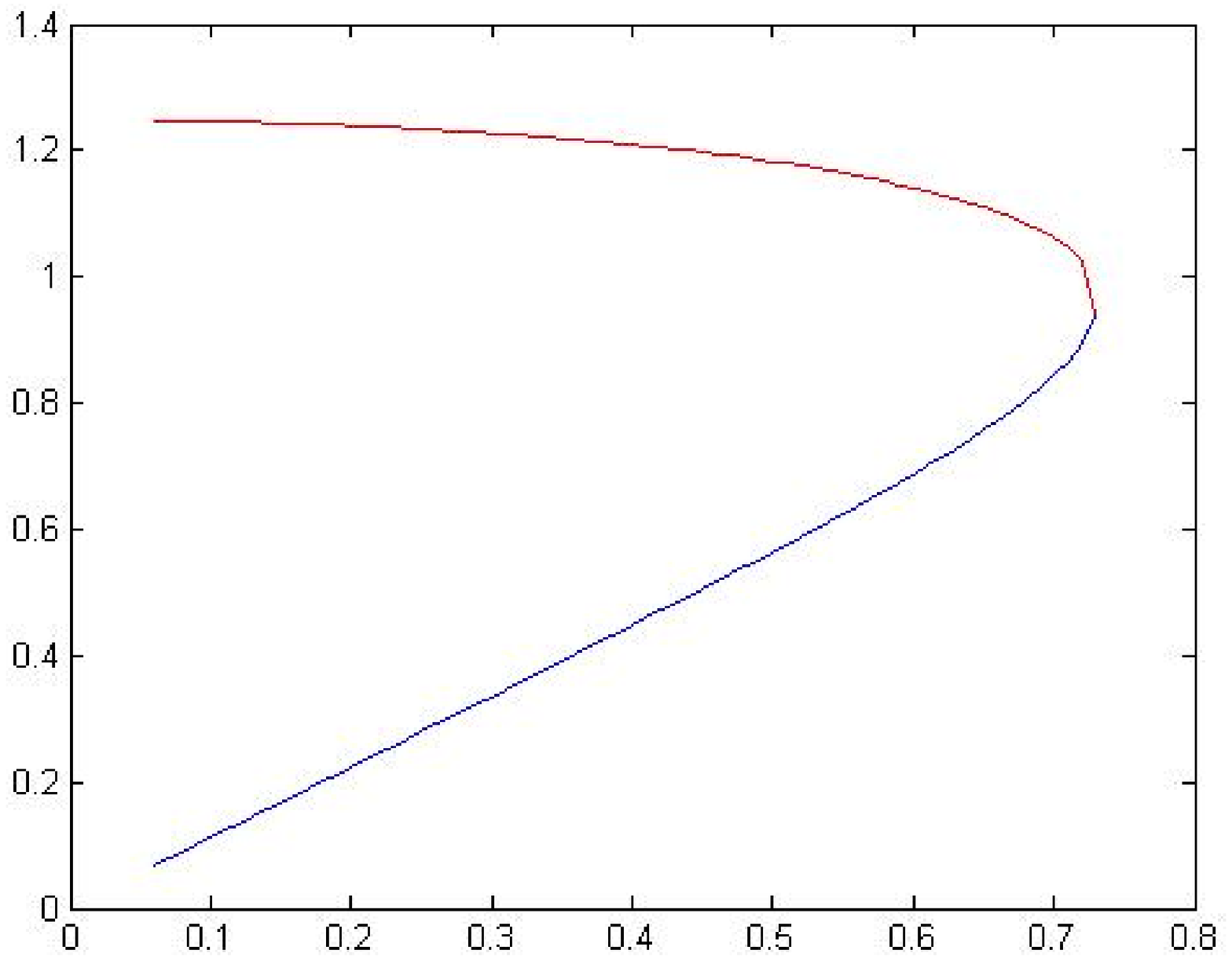}
\caption{$E$=3.4, \mbox{how $r_{1},r_{2}$ change with $L$}}\label{E3_4roots}
\end{minipage}
\begin{minipage}{0.49\textwidth}
\centering
\includegraphics[scale=0.35]{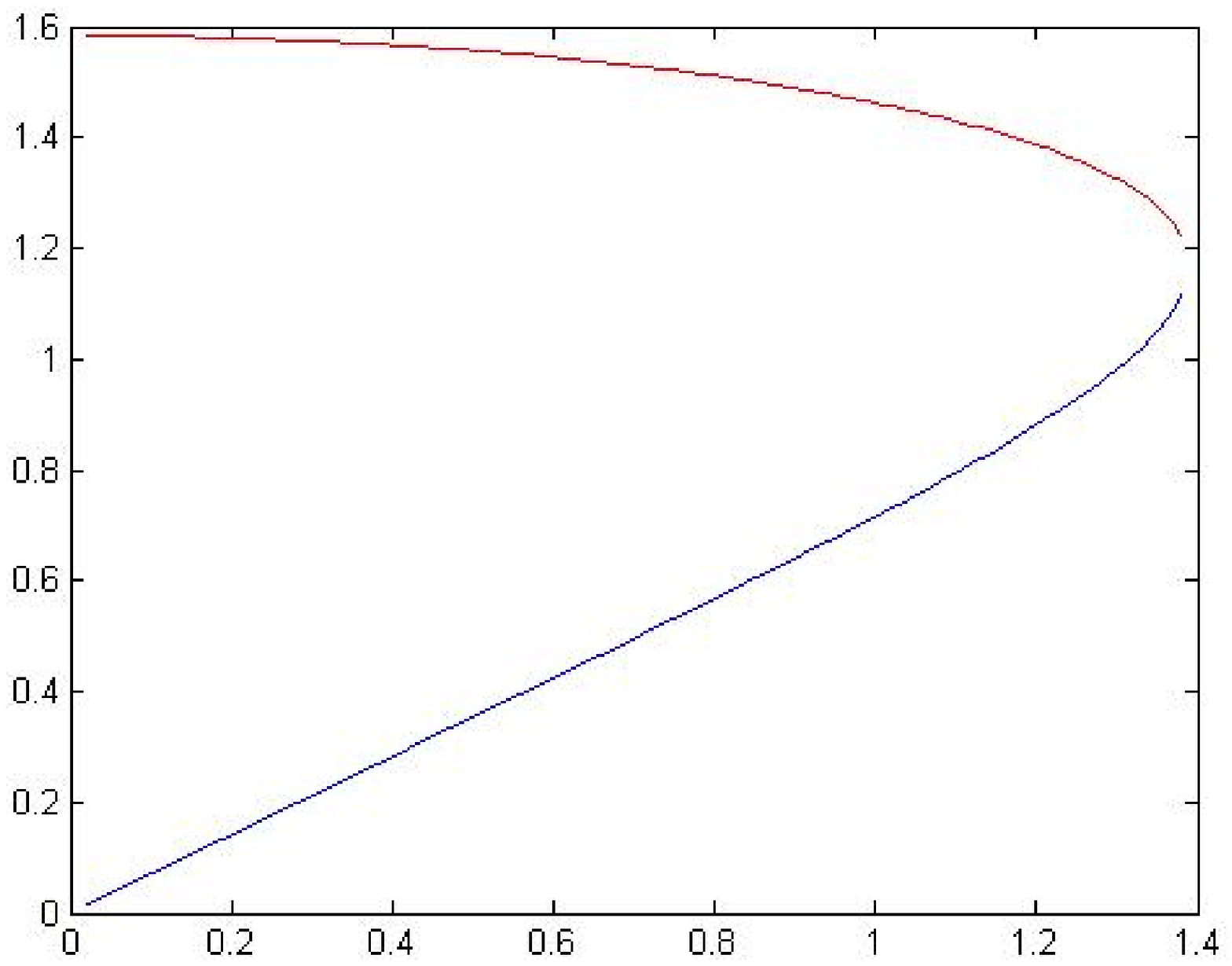}
\caption{$E$=4, \mbox{how $r_{1},r_{2}$ change with $L$}}\label{E4roots}
\end{minipage}
\end{figure}

For $L<L_{max},$ we apply Newton's iteration
\[
r_{n+1}=r_{n}-\dfrac{E-U(r)-L^{2}/2r^{2}}{-U^{\prime}(r)+L^{2}/r^{3}}%
\]
to solve (\ref{eqn-root}) with the choice of initial value
\[
r_{1,0}=\left\{
\begin{array}
[c]{l}%
0.1\quad\text{when}\quad L>0.05,\\
0.001\quad\text{when}\quad L\leq0.05
\end{array}
\right.  ,\qquad r_{2,0}=2,
\]
and the stopping criterion $|r_{n}-r_{n+1}|<10^{-10}$. We show two graphs of
$E-U(r)-L^{2}/2r^{2}$ in Figures \ref{E4_5L0_5}, and \ref{E4_8L1_2}
and the computation results are
\[
r_{1}=0.288701795314679,\ r_{2}=1.843722113067511\quad\text{for Figure
\ref{E4_5L0_5}}%
\]%
\[
r_{1}=0.634700793130700,\ r_{2}=1.938943620330099\quad\text{for Figure
\ref{E4_8L1_2}}%
\]
Figures \ref{E3_4roots} and \ref{E4roots} show how $r_{1}$ and $r_{2}$
$\left(  r_{1}<r_{2}\right)  \ $change with respect to $L$, when $E$ is given.
(Note that the equation (\ref{eqn-root}) has a unique root $r^{\ast}\ $when
$L=L_{\max}$, thus when $L$ approaches $L_{max}$, the distance $r_{2}-r_{1}%
$tends to zero.) 

\textit{Step 3. Computing }$A(\phi,\phi)$ \textit{in}
(\ref{formula-qudratic-A0}). Denote the integrand
\[
\mathcal{I}(E,L)=f_{0}^{\prime}(E,L)\int_{r1(E,L)}^{r2(E,L)}(\phi
-\overline{\phi})^{2}\frac{2Ldr}{\sqrt{2(E-U_{0}(r)-L^{2}/2r^{2})}}.
\]
Choose $\phi=e^{-r}$, Table \ref{EL} shows some of the values of
$\mathcal{I}(E,L)$. \begin{table}[ptb]
\caption{the value of $\mathcal{I}(E,L)$ for given $E$ and $L$}%
\label{EL}%
\centering
\begin{tabular}
[c]{c|c}\hline
$(E, L)$ & $intergrand(E,L)$\\\hline
(4.1,0.6) & 0.002530474111772\\\hline
(4.3,1) & 0.053619816035575\\\hline
(4.5,1.2) & -0.086444987713539\\\hline
(4.7,1.3) & -0.086591041429108\\\hline
(4.9,1.7) & -0.066267498867856\\\hline
\end{tabular}
\end{table}

For given $E$, the integration interval of $\mathcal{I}(E,L)$ in
(\ref{formula-qudratic-A0}) is $[0,L_{max}]$ for $L$. When $L=0$ and
$L=L_{max}$, (\ref{eq}) has only one solution instead of two. To avoid this
problem, we integrate $\mathcal{I}(E,L)$ for $L$ in the truncated interval
$[a_{1},L_{max}-a_{2}]$, where $a_{1},a_{2}>0$. Choose $a_{2}=10^{-5}$, and
compare the value of $\int_{a_{1}}^{L_{max}-a_{2}}\mathcal{I}(E,L)dL$ when
$a_{1}=0.01$ and $a_{1}=0.001$ in Table \ref{a_2}. We see that the error is
approximately $10^{-7}$. \begin{table}[ptbh]
\caption{Value of $\int_{a_{1}}^{L_{max}-a_{2}}\mathcal{I}(E,L)dL$ for
$a_{1}=0.01$ and $a_{1}=0.001$}%
\label{a_2}%
\centering
\begin{tabular}
[c]{c|c|c}\hline
& $a_{1}=0.01$ & $a_{1}=0.001$\\\hline
$intergrandE(4.3)$ & 0.057831884696931 & 0.057831864357174\\\hline
$intergrandE(4.5)$ & -0.084810940209437 & -0.084810952811088\\\hline
$intergrandE(4.7)$ & -0.095869703408813 & -0.095869701098749\\\hline
$intergrandE(4.9)$ & -0.058302133990229 & -0.058302146458761\\\hline
\end{tabular}
\end{table}

\textit{Step 4. Error Estimate.} Next we give a theoretical estimate for the
error introduced by the truncation. Let $A(E)=\int_{0}^{r(E)}\frac{dr}%
{\sqrt{2(E-U(r))}},$ where $r(E)$ is the unique solution of $E-U(r)=0$. For
$\phi(r)=e^{-r}$, the error term for $L$ on $[0,a_{1}]$ is
\begin{align}
&  \left\vert 32\pi^{3}\int_{4}^{5}\int_{0}^{a_{1}}\mu^{\prime}\left(
E\right)  L^{4}\int_{r_{1}(E,L)}^{r_{2}(E,L)}(\phi-\overline{\phi})^{2}%
\frac{2LdrdLdE}{\sqrt{2(E-U(r)-L^{2}/2r^{2})}}\newline\right\vert
\label{error1}\\
&  \leq\frac{64}{6}\pi^{3}\int_{4}^{5}|\mu^{\prime}(E)|\max\left\{  A\left(
E\right)  ,T\left(  E,a_{1}\right)  \right\}  dE\ a_{1}^{6}.\nonumber
\end{align}
In the above, we use the estimate
\[
T(E,L)=\int_{r_{1}(E,L)}^{r_{2}(E,L)}\frac{dr}{\sqrt{2(E-U_{0}-L^{2}/2r^{2})}%
}\leq\max\{A(E),T(E,a_{1})\}
\]
due to the facts that $\lim_{L\rightarrow0+}T\left(  E,L\right)  =A\left(
E\right)  $ and $T\left(  E,L\right)  $ is monotone for $L\ $on the small
interval $\left[  0,a_{1}\right]  $. We numerically compute the coefficient on
the right side of (\ref{error1}) by choosing $a_{1}=0.01$, then
\[
\frac{64}{6}\pi^{3}\int_{4}^{5}|\mu^{\prime}(E)|\max\left\{  A\left(
E\right)  ,T\left(  E,a_{1}\right)  \right\}  dE=3.787856674901141\times
10^{2}.
\]
Thus $a_{1}=0.01$ is enough to make the first error to be of order $10^{-10}$.

The error on $[L_{max}-a_{2},L_{max}]$ is
\begin{align}
&  \ \ \ \ \left\vert 32\pi^{3}\int_{4}^{5}\int_{L_{max}-a_{2}}^{L_{max}}%
\mu^{\prime}\left(  E\right)  L^{4}\int_{r_{1}(E,L)}^{r_{2}(E,L)}%
(\phi-\overline{\phi})^{2}\frac{2LdrdLdE}{\sqrt{2(E-U(r)-L^{2}/2r^{2})}%
}\newline\right\vert \label{error2}\\
&  \leq64\pi^{3}\int_{4}^{5}L_{max}(E)^{5}|\mu^{\prime}(E)|\max\left\{
\frac{\pi}{\sqrt{\phi_{\text{eff}}^{\prime\prime}\left(  r^{\ast};L_{\max
}\right)  }},T\left(  E,L_{\max}-a_{2}\right)  \right\}  dE\ a_{2}.\nonumber
\end{align}
where the effective potential
\[
\phi_{\text{eff}}(r)=U_{0}(r)+L^{2}/2r^{2}%
\]
Since
\[
T(E,L)\leq\max\left\{  \frac{\pi}{\sqrt{\phi_{\text{eff}}^{\prime\prime
}\left(  r^{\ast};L_{\max}\right)  }},T\left(  E,L_{\max}-a_{2}\right)
\right\}  ,\ L\in\left[  L_{\max}-a_{2},L_{\max}\right]  ,
\]
due to the facts that $\lim_{L\rightarrow L_{\max-}}T\left(  E,L\right)
=\frac{\pi}{\sqrt{\phi_{\text{eff}}^{\prime\prime}\left(  r^{\ast};L_{\max
}\right)  }}$ and $T\left(  E,L\right)  $ is monotone for$\ L\ $on $\left[
L_{\max}-a_{2},L_{\max}\right]  $.\ 

Numerically computing the coefficient before $a_{2}$ in the right hand side of
(\ref{error2}), we get $0.605275913474830$. Choose $a_{2}=10^{-5}$, the right
side of (\ref{error2}) if of order $10^{-5}$, which is good enough for what we
need later on.

\textit{Step 5. Conclusion. }We choose $\phi(r)=e^{-r}$. The first term of
(\ref{formula-qudratic-A0}) is $\int|\nabla\phi|^{2}=\pi$, and the second term
is computed to be $-31.733535998660550$, by choosing $h=0.1\ $in Runga-Kutta
Method when solving $U(r)$. Therefore $(A\phi,\phi)=\pi-31.733535998660550<0$.
We gradually decrease $h$, and repeat the whole process to calculate the
second term of (\ref{formula-qudratic-A0}). The result is shown in Table
\ref{h}. \begin{table}[ptb]
\caption{results for different $h$}%
\label{h}%
\centering
\begin{tabular}
[c]{c|c}\hline
$h$ & the second term\\\hline
0.1 & -31.733535998660550\\\hline
0.05 & -38.227746546049751\\\hline
0.025 & -41.848643231095494\\\hline
0.01 & -44.146480909452201\\\hline
0.005 & -44.933896897054382\\\hline
0.0025 & -45.331673900274509\\\hline
0.001 & -45.571655158578380\\\hline
\end{tabular}
\end{table}Thus when $h$ becomes smaller, the second term of
(\ref{formula-qudratic-A0}$)$ is less than $-45$. Combining with the error
estimates of (\ref{error1}) and (\ref{error2}) in Step 4, we can ensure
$(A_{0}\phi,\phi)<0$.

\subsection{\textbf{Numerical Computation for Example} 2}

\begin{figure}[ptb]
\centering
\includegraphics[scale=0.4]{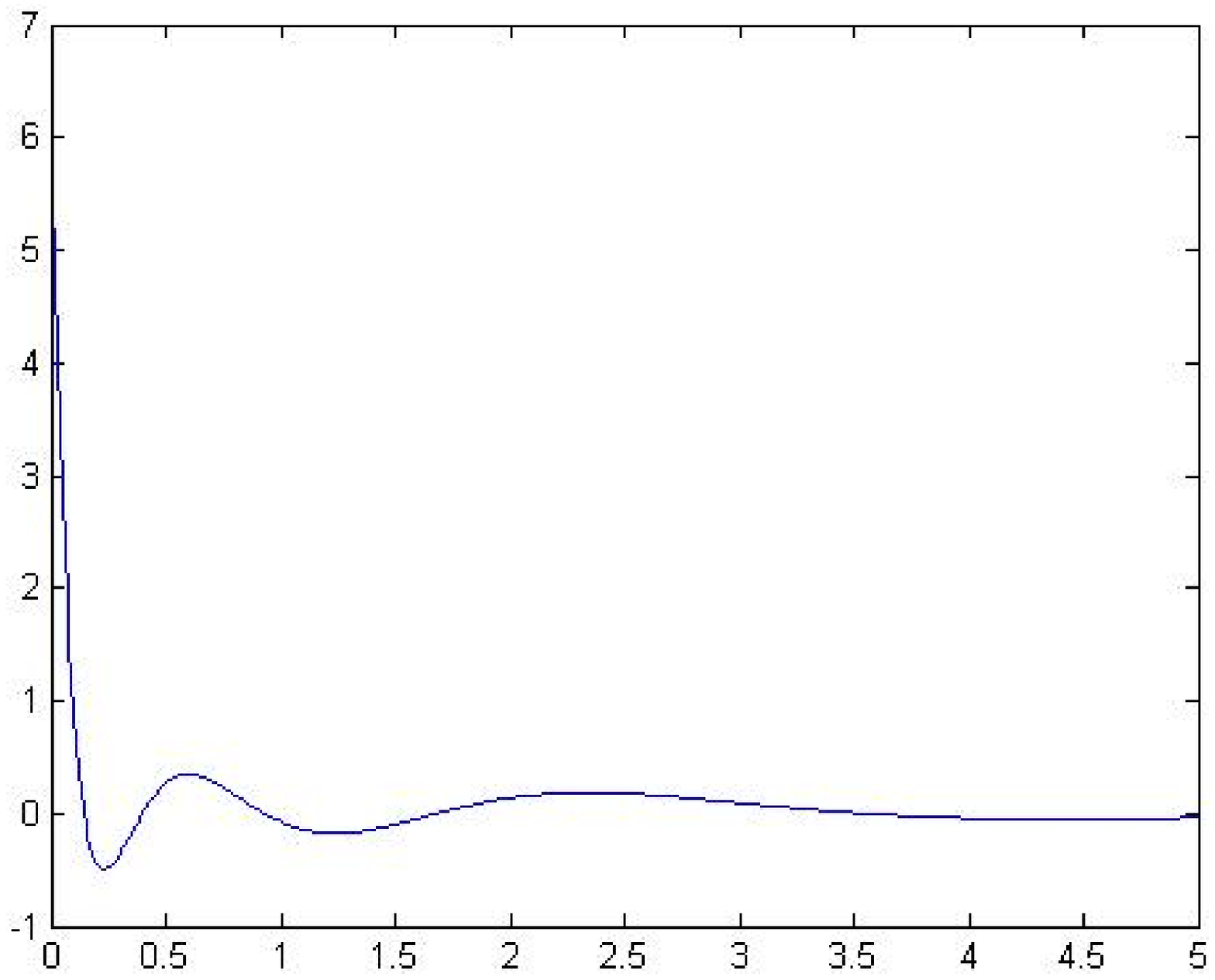}\caption{$\phi(r)$}%
\label{eigenfun}%
\end{figure}

With profile (\ref{mu2}), we choose $\phi(r)$ to be a linear combination of
several functions $e^{-\alpha_{i}r},\ 1\leq i\leq n,\ \phi(r)=\sum
a_{i}e^{-\alpha_{i}r}$, with $a_{i}$ to be determined. In the quadratic form
(\ref{formula-qudratic-A0}), the first term is computed to be
\[
\begin{aligned}\int|\nabla\phi|^{2}=4\pi(\sum\frac{a_{i}^{2}}{4\alpha_{i}%
}+\sum\frac{4\alpha_{i}\alpha_{j}}{(\alpha_{i}+\alpha_{j})^{3}}a_{i}%
a_{j})\newline\triangleq\sum b_{i}a_{i}^{2}+2\sum b_{ij}a_{i}a_{j}%
\end{aligned}
\]
and the second term is
\[
\begin{aligned}
&32\pi^{3}\int f_{0}^{\prime}(E,L)\int_{r1(E,L)}^{r2(E,L)}(\sum
a_{i}\phi_{i}-\overline{\sum a_{i}\phi_{i}})^{2}\frac{2LdrdEdL}{\sqrt
{2(E-U_{0}(r)-L^{2}/2r^{2})}}\\
&=\sum32\pi^{3}\int f_{0}^{\prime
}(E,L)\int_{r1(E,L)}^{r2(E,L)}(\phi_{i}-\overline{\phi_{i}})^{2}%
\frac{2LdrdEdL}{\sqrt{2(E-U_{0}(r)-L^{2}/2r^{2})}}a_{i}^{2}+\\
&\sum
32\pi^{3}\int f_{0}^{\prime}(E,L)\int_{r1(E,L)}^{r2(E,L)}(\phi_{i}%
-\overline{\phi_{i}})(\phi_{j}-\overline{\phi_{j}})\frac{2LdrdEdL}%
{\sqrt{2(E-U_{0}(r)-L^{2}/2r^{2})}}2a_{i}a_{j}\\
&\triangleq\sum
c_{i}a_{i}^{2}+\sum2c_{ij}a_{i}a_{j}\end{aligned}
\]
Now $(A_{0}\phi,\phi)=\sum(b_{i}+c_{i})a_{i}^{2}+2\sum(b_{ij}+c_{ij}%
)a_{i}a_{j}.$ The corresponding matrix for this quadratic form is $S=(s_{ij}%
)$, where $s_{ii}=b_{i}+c_{i},\quad s_{ij}=b_{ij}+c_{ij}$. Denote
$\lambda_{\min}$ to be the minimum eigenvalue of the matrix $S$, then
$\lambda_{\min}<0$ guarantees that there exists $\phi$ such that $(A_{0}%
\phi,\phi)<0$. We choose
\[
\phi_{1}=e^{-r},\ \phi_{2}=e^{-\frac{1}{2}r},\ \phi_{3}=e^{-\frac{1}{3}%
r},\ \phi_{4}=e^{-2r},\ \phi_{5}=e^{-3r},\ \phi_{6}=e^{-4r},\ \phi_{7}%
=e^{-5r}.\quad
\]
In (\ref{mu2}), there are five free parameters in $\mu$: $a,k,t_{1}%
,E_{0},C_{1}$ and $E_{1}$ is determined by $E_{1}=(E_{0}+t_{1}a)/(t_{1}+1)$.
Using $U_{0}$ (the initial value of the potential) as an additional parameter,
there are totally six parameters in our calculations. The idea is to view
$\lambda_{min}$ as a function of these parameters $\lambda_{min}=\lambda
_{min}(a,k,t_{1},E_{0},C_{1},U_{0}).$ Our goal now is to find proper
parameters such that $\lambda_{min}<0.$ The numerical methods are the same as
in \textit{Step 1}-\textit{Step 3} in the computation of Example 1, except
that now we use self-adaptive Runge-Kutta Method to solve (\ref{1}). We next
verify that the following choice of parameters
\[
C_{1}=2,\ E_{0}=5.1,\ a=1.9,\ k=2.01,\ t_{1}=1.5,\ U_{0}=-15.1,
\]
would lead to\ $\lambda_{min}<0\ $and$\ (A_{0}\phi,\phi)<0$. In Table
\ref{lambda} we show the results of $\lambda_{\min}$ under different
accuracies. \begin{table}[ptb]
\caption{$\lambda_{min}$ under different accuracies}%
\label{lambda}%
\centering
\begin{tabular}
[c]{c|c|c|c|c}\hline
integration accuracy & $a_{1}$ & $a_{2}$ & accuracy of the roots &
$\lambda_{min}$\\\hline
$10^{-8}$ & $10^{-5}$ & $10^{-6}$ & $10^{-10}$ & $-1.808068979998836\times
10^{-5}$\\\hline
$10^{-9}$ & $10^{-3}$ & $10^{-4}$ & $10^{-10}$ & $-3.403231868973289\times
10^{-6}$\\\hline
$10^{-9}$ & $10^{-5}$ & $10^{-6}$ & $10^{-11}$ & $-3.739973399333609\times
10^{-6}$\\\hline
$10^{-10}$ & $10^{-5}$ & $10^{-6}$ & $10^{-11}$ & $-3.196606813186484\times
10^{-6}$\\\hline
$10^{-11}$ & $10^{-5}$ & $10^{-6}$ & $10^{-11}$ & $-3.025099283901562\times
10^{-6}$\\\hline
$10^{-12}$ & $10^{-6}$ & $10^{-7}$ & $10^{-11}$ & $-3.087487453314562\times
10^{-6}$\\\hline
$10^{-13}$ & $10^{-6}$ & $10^{-7}$ & $10^{-11}$ & $-3.080483943121587\times
10^{-6}$\\\hline
$10^{-13}$ & $10^{-6}$ & $10^{-7}$ & $10^{-11}$ & $-3.089704058599446\times
10^{-6}$\\\hline
$10^{-14}$ & $10^{-7}$ & $10^{-7}$ & $10^{-11}$ & $-3.089187445892979\times
10^{-6}$\\\hline
\end{tabular}
\end{table}where $a_{1}$ and $a_{2}$ are the parameters used to truncate
$[0,L_{\max}]$ to $[a_{1},L_{\max}-a_{2}]$. We also calculate the eigenvector
corresponding to $\lambda_{\min}\ $by $\mathbf{\emph{v}}=(v_{1},v_{2}%
,v_{3},v_{4},v_{5},v_{6},v_{7})^{T}$,\ where
\begin{equation}
\begin{aligned} &v_{1}=41.702767064740\\ &v_{2}=-14.949618378683\\ &v_{3}=4.856846351504\\ &v_{4}=-201.361293458803\\ &v_{5}=571.694416252419\\ &v_{6}=-723.292461038838\\ &v_{7}=327.842857021134. \end{aligned} \label{vector-v}%
\end{equation}
Let
\[
\phi(r)=v_{1}e^{-r}+v_{2}e^{-\frac{1}{2}r}+v_{3}e^{-\frac{1}{3}r}+v_{4}%
e^{-2r}+v_{5}e^{-3r}+v_{6}e^{-4r}+v_{7}e^{-5r}.
\]
We draw a picture of this function $\phi$ in Figure \ref{eigenfun}.
Using (\ref{formula-qudratic-A0}), we obtain
\[
(A\phi,\phi)=35.231484866282720-59.247837621248109<0,
\]
when the integration accuracy is $10^{-13}$,$\ a_{1}=10^{-6}$,$\ a_{2}%
=10^{-7}$, and the root accuracy is $10^{-11}$.

\textit{Step 4. Error Estimate.} We obtain theoretical error estimate as in
Example 1. For any test function $\phi$, the error for $L$ on $[0,a_{1}]$ is
\begin{align*}
&  \left\vert 32\pi^{3}\int_{\{E|\mu^{\prime}(E)>0\}}\mu^{\prime}(E)\int
_{0}^{a_{1}}\int_{r_{1}(E,L)}^{r_{2}(E,L)}(\phi-\overline{\phi})^{2}%
\frac{2LdrdLdE}{\sqrt{2(E-U_{0}-L^{2}/2r^{2})}}\right\vert \\
\leq &  32\pi^{3}|\max\phi-\min\phi|^{2}a_{1}^{2}\int_{\{E|\mu^{\prime
}(E)>0\}}\mu^{\prime}(E)\max\{A(E),T(E,a_{1})\}dE
\end{align*}
Numerically compute the right hand side, and note
\[
|\max\phi-\min\phi|\leq2\max|\phi|\leq2\sum|v_{i}|=3.7714\times10^{3}.
\]
Choose $a_{1}=10^{-5}$, the error is no more than
\[
4.322\times10^{2}\times(3.7714\times10^{3})^{2}\times(10^{-5})^{2}%
=6.1474\times10^{-1}%
\]
Next we estimate the error for $L$ on $[L_{\max}-a_{2},L_{\max}]$. When
$r\in\lbrack r_{1},r_{2}]$, we use $|\phi-\overline{\phi}|\leq\max
|\phi^{\prime}|(r_{2}-r_{1})$ by the Mean Value Theorem. So the error is
\begin{align*}
&  32\pi^{3}\int_{\{E|\mu^{\prime}(E)>0\}}\int_{L_{max}-a_{2}}^{L_{max}}%
\int_{r_{1}(E,L)}^{r_{2}(E,L)}\mu^{\prime}(E)(\phi-\overline{\phi})^{2}%
\frac{2LdrdLdE}{\sqrt{2(E-U_{0}-L^{2}/2r^{2})}}\\
&  \leq64\pi^{3}a_{2}\left\vert \phi^{\prime}\right\vert _{L^{\infty}}^{2}%
\int_{\left\{  E|\ \mu^{\prime}\left(  E\right)  >0\right\}  }\mu^{\prime
}\left(  E\right)  L_{\max}\left(  E\right)  \left(  r_{2}(E,L_{\max}%
-a_{2})-r_{1}(E,L_{\max}-a_{2})\right)  ^{2}\cdot\\
&  \max\left\{  \frac{\pi}{\sqrt{\phi_{\text{eff}}^{\prime\prime}\left(
r^{\ast};L_{\max}\right)  }},T\left(  E,L_{\max}-a_{2}\right)  \right\}  dE\ .
\end{align*}
Note that
\[
\max\ |\phi^{\prime}|\leq5\sum|v_{i}|=9.4285\times10^{3}.
\]
Numerically computation of this error with $a_{2}=10^{-7}$ yields a bound of
\[
1.41722\times10^{-13}\times(9.4285\times10^{3})^{2}=1.3362\times10^{-6}%
\]
So the total error is at most of order $10^{-1}$, which guarantees
$(A\phi,\phi)<0.$

\section{\textbf{Summary}}

We study the instability of spherical galaxy models in the Vlasov theory for
collisionless stars. Based on the instability criterion of Theorem
\ref{theorem-insta} and careful numerical computations, we have constructed
two explicit \textit{unstable} galaxy models $f_{0}(E,L)$ and $f_{0}(E)$ with
distributions (\ref{mu1}) and (\ref{mu2}) respectively. In particular,$\ f_{0}%
(E)$ in Example 2 provides the first example of unstable isotropic galaxy
which has not been found in literature. The instability in these examples are
radial and non-oscillatory. Compared with the usual N-body codes of finding
instability of galaxy models, our method only requires numerical evaluation of
certain explicit integrals. Therefore, it is much more reliable and easier to
implement. It is hoped that Theorem \ref{theorem-insta} can be employed to
detest instability for other galaxy models in the future.

\begin{center}
\noindent{\Large Acknowledgements\ }
\end{center}

\vskip0.2cm This research is supported partly by NSF grants DMS-0603815 and
DMS-0505460 (Guo) and DMS-0908175 (Lin).


\begin{thebibliography}{99}                                                                                               %


\bibitem {an61}Antonov, V. A. \textit{Remarks on the problem of stability in
stellar dynamics.} Soviet Astr, AJ., \textbf{4}, 859-867 (1961).

\bibitem {ant62}Antonov, V. A., \textit{Solution of the problem of stability
of stellar system Emden's density law and the spherical distribution of
velocities}, Vestnik Leningradskogo Universiteta, Leningrad University, 1962.

\bibitem {bgh86}Barnes, J.; Hut, P.; Goodman, J., Dynamical instabilities in
spherical stellar systems, Astrophysical Journal, vol. \textbf{300}, p.
112-131, 1986.

\bibitem {bar71}Bartholomew, P., \textit{On the theory of stability of
galaxies}, Monthly Notices of the Royal Astronomical Society, Vol.
\textbf{151}, p. 333 (1971).

\bibitem {bertin}Bertin, Giuseppe, Dynamics of Galaxies, Cambridge University
Press, 2000.

\bibitem {BT}Binney, J., Tremaine, S., \textit{Galactic Dynamics (2nd
edition).} Princeton University Press, 2008.

\bibitem {dbf73}Doremus, J. P.; Feix, M. R.; Baumann, G.; , \textit{Stability
of Encounterless Spherical Stellar Systems}, Phys. Rev. Letts, Vol.
\textbf{26}, p. 725 (1971).

\bibitem {dbf76}Gillon, D.; Cantus, M.; Doremus, J. P.; Baumann, G.,
\textit{Stability of self-gravitating spherical systems in which phase space
density is a function of energy and angular momentum, for spherical
perturbations}, Astronomy and Astrophysics, vol. \textbf{50}, no. 3, p.
467-470, 1976.

\bibitem {fp84}Fridman, A., Polyachenko, V., \textit{Physics of Gravitating
System Vol I and II, }Springer-Verlag, 1984.

\bibitem {goodman88}Goodman, Jeremy,\textit{ An instability test for
nonrotating galaxies}, Astrophysical Journal, vol. \textbf{329}, p. 612-617, 1988.

\bibitem {GL}Guo, Y., Z. Lin, \textit{Unstable and stable galaxy models.
}Commun. Math. Phys\textit{.} \textbf{279}, no. 3, 789--813, 2008.

\bibitem {henon73}Henon, M., \textit{Numerical Experiments on the Stability of
Spherical Stellar Systems}, Astronomy and Astrophysics, Vol. 24, p. 229 (1973).

\bibitem {gr-king}Guo, Y., Rein, G., \textit{A non-variational approach to
nonlinear stability in stellar dynamics applied to the King model}, Comm.
Math. Phys. 271 (2007), no. 2, 489-509.

\bibitem {KS85}Kandrup, H.; Signet, J. F.; \textit{A simple proof of dynamical
stability for a class of spherical clusters. }The Astrophys. J. \textbf{298},
p. 27-33.(1985)

\bibitem {kandrup91}Kandrup, Henry E., \textit{A stability criterion for any
collisionless stellar equilibrium and some concrete applications thereof},
Astrophysical Journal, vol. \textbf{370}, p. 312-317, 1991.

\bibitem {king}King, Ivan R., \textit{The structure of star clusters. III.
Some simple dynamical models}, Astronomical Journal, Vol. \textbf{71}, p. 64 (1966).

\bibitem {merritt}Merritt, David, \textit{Elliptical Galaxy Dynamics}, The
Publications of the Astronomical Society of the Pacific, Volume \textbf{111},
Issue 756, pp. 129-168.

\bibitem {palmer}Palmer, P. L., \textit{Stability of collisionless stellar
systems: mechanisms for the dynamical structure of galaxies}, Kluwer Academic
Publishers, 1994.

\bibitem {aly96}Perez, Jerome and Aly, Jean-Jacques, \textit{Stability of
spherical stellar systems - I. Analytical results}, Monthly Notices of the
Royal Astronomical Society, Volume \textbf{280}, Issue 3, pp. 689-699, 1996.

\bibitem {sygnet84}Sygnet, J. F.; des Forets, G.; Lachieze-Rey, M.; Pellat,
R., \textit{Stability of gravitational systems and gravothermal catastrophe in
astrophysics}, Astrophysical Journal, vol. \textbf{276}, p. 737-745, 1984.

\bibitem {1}Rein,Gerhard; Rendall,Alan D. \textit{Compact support of
spherically symmetric equilibria in non-relativistic and relativistic galactic
dynamics,} Math. Proc. Camb. Phil. Soc. \textbf{128} (2000), 363-380.
\end{thebibliography}
\end{document}